\documentclass[UTF8,
reprint,
superscriptaddress,
showpacs,
amsmath,amssymb,
aps,
prc,
floatfix]%
{revtex4-2}

\usepackage{color}
\usepackage{graphicx}
\usepackage{dcolumn}
\usepackage{bm}
\usepackage{epstopdf}
\usepackage{multirow}
\usepackage{booktabs}
\usepackage[
bookmarks=true,colorlinks,%
citecolor=blue,linkcolor=blue,anchorcolor=blue,filecolor=blue,urlcolor=blue,%
]{hyperref}

\begin{document}

\hyphenpenalty=20000
\tolerance=2000

\title{New effective interactions for hypernuclei in density dependent relativistic mean field model}

\author{Yu-Ting Rong}

\affiliation{CAS Key Laboratory of Theoretical Physics, Institute of Theoretical Physics, 
	Chinese Academy of Sciences, Beijing, 100190, China}
\affiliation{School of Physical Sciences, University of Chinese Academy of Sciences, Beijing, 100049, China}

\author{Zhong-Hao Tu}
\affiliation{CAS Key Laboratory of Theoretical Physics, Institute of Theoretical Physics, 
	Chinese Academy of Sciences, Beijing, 100190, China}
\affiliation{School of Physical Sciences, University of Chinese Academy of Sciences, Beijing, 100049, China}

\author{Shan-Gui Zhou}
\email{sgzhou@itp.ac.cn}
\affiliation{CAS Key Laboratory of Theoretical Physics, Institute of Theoretical Physics, 
	Chinese Academy of Sciences, Beijing, 100190, China}
\affiliation{School of Physical Sciences, University of Chinese Academy of Sciences, Beijing, 100049, China}
\affiliation{Center of Theoretical Nuclear Physics, National Laboratory of Heavy Ion Accelerator, Lanzhou, 730000, China}
\affiliation{Synergetic Innovation Center for Quantum Effects and Application, Hunan Normal University, Changsha, 410081, China}

\date{\today}

\begin{abstract}
New effective $\Lambda N$ interactions are proposed for the density dependent 
relativistic mean field model. 
The multidimensionally constrained relativistic mean field model is used to 
calculate ground state properties of eleven known $\Lambda$ hypernuclei 
with $A\ge 12$ and the corresponding core nuclei.
Based on effective $NN$ interactions DD-ME2 and PKDD, the ratios $R_\sigma$ 
and $R_\omega$ of scalar and vector coupling constants between $\Lambda N$ and 
$NN$ interactions are determined by fitting calculated $\Lambda$ separation 
energies to experimental values. 
We propose six new effective interactions for $\Lambda$ hypernuclei: 
DD-ME2-Y1, DD-ME2-Y2, DD-ME2-Y3, PKDD-Y1, PKDD-Y2 and PKDD-Y3
with three ways of grouping and including these eleven hypernuclei in the fitting. 
It is found that the two ratios $R_\sigma$ and $R_\omega$ are correlated well 
and there holds a good linear relation between them. 
The statistical errors of the ratio parameters in these effective interactions are analyzed.
These new effective interactions are used to study the equation of state of
hypernuclear matter and neutron star properties with hyperons.
\end{abstract}

\maketitle










\section{Introduction}\label{sec1}

Hyperon-nucleon ($YN$) interaction is an important part of baryon-baryon interactions~\cite{Epelbaum2009_RMP81-1773,Rijken2010_PTPSupp185-14,Machleidt2016_PS91-083007,
Ren2018_ChinPhysC42-14103,Li2018_ChinPhysC42-014105,Laehde2019_LatticeEFT}.
It is crucial for understanding the structure of hypernuclei and properties 
of compact stars and vice versa, i.e., the studies of hypernuclear structure 
and compact stars are crucial for constraining the $YN$ interaction~\cite{Gal2016_RMP88-035004,Tolos2020_PPNP112-103770}.
Many experiments have been carried out to study hypernuclear structure at 
LHC, DA$\Phi$NE, JLab, J-PARC, MAMI, RHIC and GSI~\cite{Tamura2005_NPA754-58,Hashimoto2006_PPNP57-564,Tamura2012_PTEP2012-02B012,
Esser2013_NPA914-519,Ma2013_JPCS420-12036,Feliciello2015_RPP78-096301}. 
The investigations of single-$\Lambda$, double-$\Lambda$~\cite{Takahashi2001_PRL87-212502}, 
$\Xi$ hypernuclei~\cite{Nakazawa2015_PTEP2015-033D02,Hayakawa2021_PRL126-062501} and 
anti-hypernuclei~\cite{STAR2020_NatPhys16-409} have provided valuable information 
on $YN$ and $YY$ interactions. 
From the theory side, the SU(3) flavor symmetry, which reveals the relation 
between baryon and meson octet states, is generally used for determining 
the $YN$ interaction. 
With the constraint of SU(3) flavor symmetry or experimental observables, 
many approaches, such as Lattice QCD~\cite{Beane2011_PPNP66-1,Beane2013_PRC88-024003,Aoki2012_PTEP2012-01A105,
Sasaki2015_PTEP2015-113B01}, 
the chiral effective field theory~\cite{Polinder2006_NPA779-244,Haidenbauer2020_EPJA56-55,Ren2020_PRC101-034001}, 
the Nijmegen soft-core model~\cite{Rijken1999_PRC59-21,Rijken2010_PTPSupp185-14,Nagels2019_PRC99-044003}, 
the J\"{u}lich hyperon-nucleon model~\cite{Holzenkamp1989_NPA500-485,Reuber1996_NPA608-243,Haidenbauer2005_PRC72-044005}, 
the Skyrme Hartree-Fock model~\cite{Lanskoy1998_PRC58-3351,Guleria2012_NPA886-71,Schulze2014_PRC90-047301} and 
the relativistic mean field (RMF) model~\cite{Schaffner1994_AP235-35,Mares1994_PRC49-2472,Shen2006_PTP115-325,
Rong2020_PLB807-135533}, 
have been used to study the structure of hypernuclei based on various 
(effective) $YN$ and $YY$ interactions.

In the RMF model proposed by Walecka~\cite{Walecka1974_APNY83-491}, 
there are only linear coupling terms in the interaction Lagrangian,
i.e., mesons do not interact among themselves.
Such linear couplings lead to an improper description of the 
incompressibility of nuclear matter and surface properties in finite nuclei.
To describe adequately these essential nuclear properties, 
either nonlinear (NL) self-couplings of meson fields~\cite{Boguta1977_NPA292-413,Serot1979_PLB86-146,Reinhard1988_ZPA329-257,
Sugahara1994_NPA579-557,Long2004_PRC69-034319}
or density dependent (DD) nucleon-meson couplings~\cite{Brockmann1992_PRL68-3408,Haddad1993_PRC48-2740,Boersma1994_PRC49-233,
Fuchs1995_PRC52-3043,Lenske1995_PLB345-355,Typel1999_NPA656-331,Niksic2002_PRC66-024306}
have been introduced in the RMF model.
The NL-RMF model has been extended and widely applied to the study of hypernuclei 
and many effective interactions have been proposed.
However, there are much less DD-RMF effective interactions for hypernuclei~\cite{Keil2000_PRC61-064309,Hofmann2001_PRC64-025804,Finelli2009_NPA831-163,
Colucci2013_PRC87-055806,Banik2014_ApJSupp214-22,van-Dalen2014_PLB734-383,
Fortin2017_PRC95-065803,Providencia2019_FASS6-13}.
The effective $\Lambda N$ interactions were usually obtained by fitting the calculated 
$\Lambda$ separation energies to experimental values of known hypernuclei and 
in the fitting procedure the RMF calculations were carried out with 
the restriction of spherical symmetry.
Yet most observed hypernuclei are deformed and there are only few spherical hypernuclei. 
In this letter, we propose new DD effective interactions for hypernuclei by using 
a deformed RMF model---the multidimensionally constrained (MDC) RMF model. 

The paper is organized as follows. 
In Sec.~\ref{sec2}, we introduce the MDC-RMF model for $\Lambda$ hypernuclei 
with density dependent couplings. 
In Sec.~\ref{sec3}, we present our results and discuss $\Lambda$ separation
energies, deformation effects, parameter correlations and neutron star properties 
with the new effective interactions.
Finally a summary is given in Sec.~\ref{sec4}.

\section{Theoretical framework}\label{sec2}

The RMF model has been very successful in describing properties of nuclear matter
and finite nuclei in the whole nuclear chart~\cite{Serot1986_ANP16-1,Reinhard1989_RPP52-439,
Ring1996_PPNP37-193,Bender2003_RMP75-121,Vretenar2005_PR409-101,
Meng2006_PPNP57-470,Niksic2011_PPNP66-519,Liang2015_PR570-1,
Meng2015_JPG42-093101,Zhou2016_PS91-063008,Meng2016_RDFNS}. 
Both the NL-RMF and DD-RMF models have been extended to the study of hypernuclei~\cite{Rufa1990_PRC42-2469,Glendenning1991_PRL67-2414,Schaffner1994_AP235-35,
Mares1994_PRC49-2472,Ma1996_NPA608-305,Vretenar1998_PRC57-R1060,Shen2006_PTP115-325,
Keil2000_PRC61-064309,Hofmann2001_PRC64-025804,Finelli2009_NPA831-163,
Wang2010_PRC81-025801,Lu2011_PRC84-014328,Tanimura2012_PRC85-014306,
Colucci2013_PRC87-055806,van-Dalen2014_PLB734-383,Lu2014_PRC89-044307,
Banik2014_ApJSupp214-22,Ren2017_PRC95-054318,Fortin2017_PRC95-065803,
Sun2017_PRC96-044312,Providencia2019_FASS6-13,Rong2020_PLB807-135533}.
In the DD-RMF model, the Lagrangian for a $\Lambda$ hypernucleus is written as
\begin{widetext}
\begin{equation}\label{eq:Lagrangian}
\begin{split}
{\cal L} = & \sum_{B}\bar{\psi}_B
\left(
i \gamma_\mu \partial^\mu - M_B 
- g_{\sigma B} \sigma - g_{\sigma^* B} \sigma^*
- g_{\omega B} \gamma_\mu \omega^\mu 
- g_{\phi B}   \gamma_\mu \phi^\mu 
- g_{\rho B}   \gamma_\mu \vec{\tau} \cdot \vec{\rho}^\mu         
- e \gamma_\mu \dfrac{1-\tau_3}{2} A^\mu
\right) \psi_B 
\\ &
+ \psi_\Lambda\dfrac{f_{\omega\Lambda\Lambda}}{4M_\Lambda} \sigma_{\mu\nu} \Omega^{\mu\nu}\psi_\Lambda
+ \dfrac{1}{2} \partial^\mu \sigma   \partial_\mu \sigma   - \dfrac{1}{2} m_\sigma^2   \sigma^2            
+ \dfrac{1}{2} \partial_\mu \sigma^* \partial^\mu \sigma^* - \dfrac{1}{2} m_{\sigma^*} \sigma^{*2}
- \dfrac{1}{4} \Omega^{\mu\nu} \Omega_{\mu\nu} + \dfrac{1}{2} m_\omega^2 \omega^\mu \omega_\mu
\\ & 
- \dfrac{1}{4}      S^{\mu\nu}      S_{\mu\nu}   + \dfrac{1}{2} m_\phi^2   \phi^\mu   \phi_\mu
- \dfrac{1}{4} \vec{R}^{\mu\nu} \vec{R}_{\mu\nu} + \dfrac{1}{2} m_\rho^2 \vec{\rho}^\mu \vec{\rho}_\mu 
- \dfrac{1}{4} F^{\mu\nu}F_{\mu\nu}         
,
\end{split}
\end{equation}
\end{widetext}
where $B$ represents neutron, proton or $\Lambda$ and $M_B$ is the corresponding mass. 
$\sigma$ and $\sigma^*$ are scalar-isoscalar meson fields coupled to baryons, 
$\omega^\mu$ and $\phi^\mu$ are vector-isoscalar meson fields coupled to baryons, 
$\vec{\rho}^\mu$ is the vector-isovector meson field coupled to nucleons and 
$A^\mu$ is the photon field.  
$\Omega_{\mu\nu}$, $S_{\mu\nu}$, $\vec{R}_{\mu\nu}$ and $F_{\mu\nu}$ are field 
tensors of the vector mesons $\omega^\mu$, $\phi^\mu$ and $\vec{\rho}^\mu$ 
and photons $A^\mu$. 
$m_\sigma$ ($g_{\sigma B}$), $m_{\sigma^*}$ ($g_{\sigma^* B}$), 
$m_\omega$ ($g_{\omega B}$), $m_\phi$ ($g_{\phi B}$) and $m_\rho$ ($g_{\rho B}$) 
are the masses (coupling constants) for meson fields. 
The coupling constants are dependent on the total baryonic density $\rho^{\upsilon}$,
\begin{equation} 
 g_{mB}(\rho^\upsilon) = g_{mB}(\rho_{\text{sat}}) f_{mB}(x),\ \ \ 
 x = \rho^{\upsilon}/\rho_{\text{sat}},
 \label{eq:DD-rho0}
\end{equation}
where $m$ represents mesons and $\rho_\text{sat}$ is the saturation density of 
nuclear matter. 
$g_{mB}(\rho_{\mathrm{sat}})$, the coupling constant at saturation density and $f_{mB}(x)$ describing the density dependence will be discussed in Sec.~\ref{sec3}.

Starting for the Lagrangian (\ref{eq:Lagrangian}), the equations of motion 
can be derived via the variational principle.
The Dirac equation for baryons reads 

\begin{eqnarray}
&  \left[ \bm{\alpha} \cdot \bm{p} + V_B + T_B + \Sigma_R + \beta(M_B + S_B) 
 \right]       \psi_{iB}
  =  
 \varepsilon_i \psi_{iB} ,
\label{eq:Dirac}
\end{eqnarray}
 the Klein-Gordon equations for mesons and the Proca equation for photon are
\begin{equation}
\begin{split}
 (-\Delta + m_\sigma^2)     \sigma   & = -g_{\sigma N}         \rho_N^s        - g_{\sigma\Lambda} \rho_\Lambda^s, \\
 (-\Delta + m_{\sigma^*}^2) \sigma^* & = -g_{\sigma^* \Lambda} \rho_\Lambda^s, \\
 (-\Delta + m_\omega^2)     \omega_0 & =  g_{\omega N}         \rho_N^\upsilon + g_{\omega\Lambda} \rho_\Lambda^\upsilon    
                                         -\dfrac{f_{\omega\Lambda\Lambda}}{2M_\Lambda} \rho_\Lambda^T, \\
 (-\Delta \phi + m_\phi^2)  \phi     & =  g_{\phi\Lambda}      \rho_\Lambda^\upsilon, \\
 (-\Delta + m_\rho^2)       \rho_0   & =  g_{\rho N}          (\rho_n^\upsilon - \rho_p^\upsilon), \\
  -\Delta A_0                        & =  e                    \rho_p^\upsilon.
\end{split}
\label{eq:KG}
\end{equation}
The eq.~(\ref{eq:Dirac}) and eqs.~(\ref{eq:KG}) 
are coupled via the scalar, vector and tensor densities
\begin{equation}\label{eq:densities}
\begin{split} 
 \rho_B^s        & =  \sum_i \bar{\psi}_{iB}          \psi_{iB}, \\ 
 \rho_B^\upsilon & =  \sum_i \bar{\psi}_{iB} \gamma^0 \psi_{iB}, \\ 
 \rho_\Lambda^T  & = i \bm{\partial} 
                     \left(
                      \sum_i \psi_{i\Lambda}^\dag \bm{\gamma} \psi_{i\Lambda}
                     \right),
\end{split}
\end{equation}
and various potentials
\begin{equation}\label{eq:potentials}
\begin{split}
 V_B        = &~  g_{\omega B}\omega_0         +  g_{\phi B}\phi_0        
                + g_{\rho B}\tau_3 \rho_0      + e\dfrac{1-\tau_3}{2}A_0,\\
 S_B        = &~  g_{\sigma B}\sigma           +  g_{\sigma^* B}\sigma^*, \\
 T_\Lambda  = &~- \dfrac{ f_{\omega\Lambda\Lambda} }{ 2M_\Lambda } \beta(\bm{\alpha}\cdot \bm{p}) \omega_0, \\
 \Sigma_R   = &~  \dfrac{ \partial g_{\sigma N} }           { \partial \rho^{\upsilon} }                              
                  \rho_N^s                          \sigma
                + \dfrac{ \partial g_{\omega N} }            { \partial \rho^{\upsilon} } 
                \rho_N^\upsilon                     \omega_0
                + \dfrac{ \partial g_{\rho N} }              { \partial \rho^{\upsilon} }          
                ( \rho_n^\upsilon - \rho_p^\upsilon ) \rho_0  \\
                &
                + \dfrac{ \partial g_{\sigma^*\Lambda} }     { \partial \rho^{\upsilon} }            \rho_\Lambda^s                      \sigma^* 
                + \dfrac{ \partial g_{\phi\Lambda} }         { \partial \rho^{\upsilon} }            \rho_\Lambda^\upsilon               \phi_0 \\
                &+ \dfrac{ 1 }{ 2M_\Lambda } 
                  \dfrac{ \partial f_{\omega\Lambda\Lambda} }{ \partial \rho^{\upsilon} } 
                  \rho_\Lambda^T \omega_0.
\end{split}
\end{equation}
The rearrangement term $\Sigma_R$ is present in the DD-RMF model 
to ensure the energy-momentum conservation and thermodynamic consistency~\cite{Lenske1995_PLB345-355}.

Under the mean field and no-sea approximations, 
the Dirac equation (\ref{eq:Dirac}), the Klein-Gordon equations and the Proca equation~(\ref{eq:KG}) 
have been solved in different bases, including the coordinate space~\cite{Horowitz1981_NPA368-503,Price1987_PRC36-354,Poeschl1997_CPC101-75,
Poschl1997_CPC103-217}, the harmonic oscillator basis~\cite{Gambhir1990_APNY198-132,Stoitsov1998_PRC58-2092,Geng2007_CPL24-1865},
the Woods-Saxon basis~\cite{Zhou2003_PRC68-034323}
and the Lagrange mesh~\cite{Typel2018_FPhys6-73}.
To simplify the solving procedure of these equations, most RMF models were 
developed with certain spatial symmetries imposed on nuclei.
Note that the solution of the Dirac equation in a 3D lattice was achieved
recently and in such a RMF model, the nuclei in question are not restricted by 
any spatial symmetry~\cite{Ren2017_PRC95-024313,Ren2019_SciChinaPMA62-112062}.
In the present work, we use the MDC-RMF model~\cite{Lu2012_PRC85-011301R,Lu2014_PRC89-014323,Zhou2016_PS91-063008} 
in which an axially deformed harmonic oscillator (ADHO) basis~\cite{Gambhir1990_APNY198-132} is used,
the pairing correlations are treated by the BCS approach, and 
the $V_4$ symmetry is assumed for nuclear shapes, i.e., 
all deformations characterized by $\beta_{\lambda\mu}$ with even $\mu$, 
e.g., $\beta_{20}$, $\beta_{22}$, $\beta_{30}$, $\beta_{32}$, $\beta_{40}$, 
$\cdots$, are included self-consistently.
The MDC-RMF model has been used to study 
the potential energy surfaces and fission barriers of heavy and superheavy nuclei~\cite{Lu2012_PRC85-011301R,Lu2014_PRC89-014323,Lu2014_PS89-054028,
Zhao2015_PRC91-014321,Meng2020_SciChinaPMA63-212011}, 
non-axial octupole $Y_{32}$ correlations~\cite{Zhao2012_PRC86-057304,Zhao2017_PRC95-014320}
and the octupole correlations in chiral doublet bands~\cite{Liu2016_PRL116-112501,Chen2016_PRC94-021301R}, etc.

Both NL and DD couplings have been implemented in the MDC-RMF model for normal nuclei.
The MDC-RMF model with NL self-couplings of meson fields has been also extended 
to $\Lambda$ hypernuclei~\cite{Lu2011_PRC84-014328,Lu2014_PRC89-044307,Rong2020_PLB807-135533}.
In the present work, we have included the DD couplings in the MDC-RMF model
for $\Lambda$ hypernuclei and, under the $V_4$ symmetry,
solved Eqs.~(\ref{eq:Dirac}) and~(\ref{eq:KG}) in the ADHO basis.
To keep the time reversal symmetry, when dealing with the unpaired baryon, 
we adopt the equal filling approximation which has been widely used 
in mean field calculations~\cite{Perez-Martin2008_PRC78-014304}. 
The spurious motion due to the breaking of the translational invariance is 
treated by including the center of mass correction
$E_{\text{c.m.}} = -\langle P^2 \rangle/2M$ with $M=(A-1)M_N+M_\Lambda$
in the binding energy.

\section{Results and discussions}\label{sec3}

In this section we will determine the parameters of coupling constants 
$g_{mB}(\rho^\upsilon)$ in Eq.~({\ref{eq:DD-rho0}}) for hypernuclei based on 
available DD-RMF effective interactions for normal nuclei.
For the mesons in Eq.~(\ref{eq:Lagrangian}), 
one usually considers $\sigma$, $\omega$ and $\rho$ for normal nuclei. 
In this work, we focus on single-$\Lambda$ hypernuclei 
in which $\sigma^*$ and $\phi$ can be omitted because they only couple to 
strange quarks according to the OZI rule.
Furthermore, the electromagnetic field and $\rho$ mesons do not couple to 
$\Lambda$ because $\Lambda$ hyperon is charge neutral with isospin $\tau=0$.
Therefore we are left with $g_{\sigma \Lambda}$ and $g_{\omega\Lambda}$ to be fixed.
Many DD-RMF effective interactions have been proposed for normal nuclei,
e.g., 
TW99~\cite{Typel1999_NPA656-331}, 
DD-ME1~\cite{Niksic2002_PRC66-024306}, 
PKDD~\cite{Long2004_PRC69-034319}, 
DD-ME2~\cite{Lalazissis2005_PRC71-024312}, 
DD~\cite{Typel2005_PRC71-064301},
D$^3$C~\cite{Typel2005_PRC71-064301},
PKO1 and PKO2~\cite{Long2006_PLB640-150},
PKO3~\cite{Long2008_EPL82-12001},
PKA1~\cite{Long2007_PRC76-034314}, 
DD2~\cite{Typel2010_PRC81-015803},
DDME$\delta$~\cite{Roca-Maza2011_PRC84-054309},
DDME-X~\cite{Taninah2020_PLB800-135065}, 
DD-LZ1~\cite{Wei2020_ChinPhysC44-074107} and
DDV, DDS, DDVT, DDST, DDVTD and DDSTD~\cite{Typel2020_EPJA56-160}.
Most of them can provide a good description for the properties not only of
nuclear matter but also of finite nuclei around and far from the valley of 
$\beta$ stability.
In this work we choose two typical ones, 
PKDD~\cite{Long2004_PRC69-034319} and DD-ME2~\cite{Lalazissis2005_PRC71-024312},
in which the density dependence in Eq.~(\ref{eq:DD-rho0}) is taken as
\begin{equation}
 f_{mN}(x) = 
 \left\{
 \begin{array}{l}
  a_m \displaystyle\frac{1 + b_m \left(x+d_m\right)^2}{ 1 + c_m \left(x+d_m\right)^2 }, \ \ m = \sigma~ \text{or}~\omega,\\
  e^{\displaystyle -a_\rho(x-1)}, \ \ m = \rho, \\
 \end{array}
 \right.
\end{equation}
with nine parameters. Under five constraints, $f_{\sigma}(1)=1, f_{\omega}(1)=1, 
f_{\sigma}''(0)=0, f_{\omega}''(0)=0$ and  $f_\sigma''(1)=f_\omega''(1)$, 
there are four free parameters which have been adjusted to properties of 
nuclear matter and selected finite nuclei~\cite{Long2004_PRC69-034319,Lalazissis2005_PRC71-024312}.
We assume the same density dependence of $g_{m\Lambda}(\rho^{\upsilon})$ as 
that of $g_{mN}(\rho^\upsilon)$ ($m=\sigma$ or $\omega$).
So there are two parameters $g_{\sigma\Lambda}(\rho_{\text{sat}})$ and 
$g_{\omega\Lambda}(\rho_{\text{sat}})$ to be determined.

In $\Lambda$ hypernuclei, the single particle potential depth for $\Lambda$ 
is about 30 MeV~\cite{Bouyssy1976_PLB64-276,Millener1988_PRC38-2700} and 
the energy splitting between spin partners in single-$\Lambda$ states is 
very small compared to that for nucleons 
\cite{Ajimura2001_PRL86-4255,Akikawa2002_PRL88-082501,Kohri2002_PRC65-034607}. 
The shallow potential and small spin-orbit splittings for $\Lambda$ have been 
understood under the following mechanisms: 
(i) The effective scalar and vector boson exchange interactions with $\Sigma$ 
and $\Delta$-isobar intermediate states~\cite{Brockmann1977_PLB69-167};
(ii) the combined quark-gluon exchange between the valence baryon and 
the nucleons of the core~\cite{Pirner1979_PLB85-190};
(iii) a weak SU(3) symmetry breaking and a tensor $\omega\Lambda\Lambda$ coupling~\cite{Noble1980_PLB89-325,Jennings1990_PLB246-325}. 
In the RMF model, the last mechanism has been used mostly for the study of 
$\Lambda$ hypernuclei~\cite{Sugahara1994_PTP92-803,Ma1996_NPA608-305,
Lu2011_PRC84-014328,Wang2013_CTP60-479,Ren2017_PRC95-054318,Sun2017_PRC96-044312}. 
According to the OZI rule, the tensor coupling constant 
$f_{\omega\Lambda\Lambda}$ is the same as that of $g_{\omega\Lambda}$ 
and 
$R_{\omega\Lambda\Lambda} = f_{\omega\Lambda\Lambda} / g_{\omega\Lambda}= -1.0$~\cite{Jennings1990_PLB246-325,Cohen1991_PRC44-1181}.
In the present work we follow this convention concerning the tensor
coupling between $\omega$ and $\Lambda$ and assume 
$f_{\omega\Lambda\Lambda}(\rho^\upsilon)=- g_{\omega\Lambda}(\rho^\upsilon)$.
According to the quark model, the ratio 
$R_m = g_{m\Lambda}(\rho_\text{sat}) / g_{mN}(\rho_\text{sat}) = 2/3$ with 
$m=\sigma$ or $\omega$~\cite{Dover1984_PPNP12-171,Schaffner1994_AP235-35,Lim2018_PRD97-023010}.
These two ratios, together with the parameters of DD-ME2 or PKDD, 
define completely the DD-RMF functionals for $\Lambda$ hypernuclei. 
We use DD-ME2-Y0 and PKDD-Y0 to label these two effective interactions.
Since they cannot give even a decent description for $\Lambda$ separation 
energies of hypernuclei with $A \ge 12$, 
as seen in Table~\ref{tab:DD-ME_fitting_result}, 
we will adjust the ratios $R_\sigma$
and $R_\omega$ to experimental values of $\Lambda$ separation energies of 
selected hypernuclei.

Until now $\Lambda$ separation energies of 
33 single-$\Lambda$ hypernuclei have been measured. 
Most of these hypernuclei are very light and there are fifteen with $A \ge 12$: 
$^{12}_{~\Lambda}$B, $^{12-14}_{~~~~\Lambda}$C, $^{15,16}_{~~~~\Lambda}$N,
$^{16}_{~\Lambda}$O, $^{28}_{~\Lambda}$Si, $^{32}_{~\Lambda}$S, 
$^{40}_{~\Lambda}$Ca, $^{51,52}_{~~~~\Lambda}$V, $^{89}_{~\Lambda}$Y, 
$^{139}_{~~\Lambda}$La and $^{208}_{~~\Lambda}$Pb.
In the present work, the difference of $\Lambda$ separation energies of 
mirror hypernuclei, ($^{12}_{~\Lambda}$B, $^{12}_{~\Lambda}$C) and 
($^{16}_{~\Lambda}$N, $^{16}_{~\Lambda}$O), cannot be reproduced because 
$\Lambda$ interacts with both protons and neutrons via the exchange of 
the same mesons and the charge symmetry breaking is not enough. 
Therefore we select only one hypernucleus in each pair, namely, 
$^{12}_{~\Lambda}$C and $^{16}_{~\Lambda}$O.
$^{14}_{~\Lambda}$C and $^{15}_{\Lambda}$N are not included because their separation 
energies were measured only by using the photographic emulsion technique 
which should not be trusted for hypernuclei with $A \ge 12$~\cite{Gal2016_RMP88-035004}.
Up to date, all effective $YN$ interactions used in the RMF model were obtained 
with the restriction of spherical symmetry though many of the observed 
hypernuclei are deformed.
In this work, we will use the MDC-RMF model and consider axial 
and reflection symmetric deformations 
when adjusting the parameters of DD-RMF effective interactions.
For simplicity, the pairing correlations are ignored. 

\begin{figure*}[!htbp]
	\centering
	\includegraphics[width=0.8\textwidth]{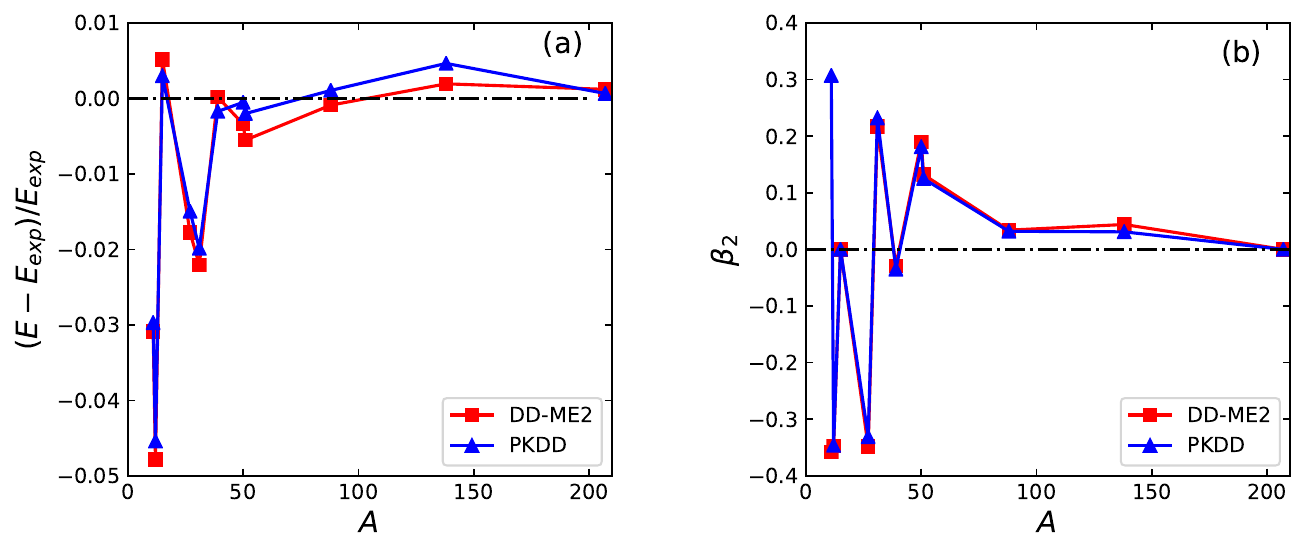}
	\caption{(Color online) 
		(a) The relative deviations of the calculated binding energies of 
		$^{11,12}$C, $^{15}$O, 
		$^{27}$Si, $^{31}$S, $^{39}$Ca, $^{50,51}$V, $^{88}$Y, $^{138}$La and $^{207}$Pb
		from the experimental values~\cite{Wang2017_ChinPhysC41-030003} and (b) the
		quadrupole deformation parameters of these nuclei.
		The MDC-RMF calculations are carried out with density dependent 
		interactions DD-ME2 and PKDD, respectively. 
	}\label{fig:NN_force_normal_nuclei}
\end{figure*}

We first calculate the ground state properties of the core nuclei 
of these eleven selected hypernuclei: $^{11,12}$C, $^{15}$O, $^{27}$Si, 
$^{31}$S, $^{39}$Ca, $^{50,51}$V, $^{88}$Y, $^{138}$La and $^{207}$Pb. 
The relative deviations of the calculated binding energies from the experimental 
values are shown in Fig.~\ref{fig:NN_force_normal_nuclei}(a). 
The experimental binding energies are taken from AME2016~\cite{Wang2017_ChinPhysC41-030003}.
From Fig.~\ref{fig:NN_force_normal_nuclei}(a),
one can see that binding energies calculated with both DD-ME2 and PKDD 
are close to the experimental values for all of these nuclei.
The relative deviations from AME2016 are within 5\% and 
the largest deviation occurs for $^{12}$C.
Except $^{15}$O and $^{207}$Pb, all the other nuclei are deformed though 
$^{39}_{~\Lambda}$Ca is almost spherical, 
as seen in Fig.~\ref{fig:NN_force_normal_nuclei}(b). 
From MDC-RMF calculations with PKDD and DD-ME2, similar deformations are 
obtained for ten of these eleven nuclei with $^{11}$C as an exception:
$^{11}$C is oblate with DD-ME2 but prolate with PKDD.

\begin{table*}[htbp]
\centering
\caption{The calculated $\Lambda$ separation energies $B_\Lambda$ (in MeV) for selected 
hypernuclei with DD-ME2-Y$i$ and PKDD-Y$i$ ($i=0,1,2$ and $3$) in comparison 
with experimental values. 
$\bar{\chi}^2$ ($\bar{\chi}^2_{\rm all}$) represents the average least-square 
deviation of the calculated $\Lambda$ separation energies of hypernuclei 
in each group (all of the eleven hypernuclei) from the experimental values. 
The bold-faced $B_\Lambda$ values denote that the experimental values
of the corresponding hypernuclei are used in the parametrization fitting.
See text for the grouping of hypernuclei used in the fitting.
The root mean square (rms) deviation $\Delta$ is given in MeV and the root of 
relative square (rrs) deviation $\delta$ is in percentage. 
The corresponding ratios $R_\sigma$ and $R_\omega$ of the effective $\Lambda N$ 
interactions together with statistical errors $\sigma_{R_\sigma}^+$ and $\sigma_{R_\sigma}^-$ of $R_{\sigma}$ are listed at the bottom of the table. 
The tensor coupling constant   
$f_{\omega\Lambda\Lambda} = - g_{\omega\Lambda}$. }
{
\setlength\tabcolsep{7pt}
	\begin{tabular}{ccrrrrrrrr}
	\hline
	\hline
                          &              & \multicolumn{4}{c}{DD-ME-Y$i$}
                                         & \multicolumn{4}{c}{PKDD-Y$i$}      \\
    Hypernuclei           &   Exp.       & $i=0$~     & $i=1$~     & $i=2$~     & $i=3$~
                                         & $i=0$~     & $i=1$~     & $i=2$~     & $i=3$~      \\
    \hline
    $^{12}_{~\Lambda}$C   &$11.36\pm0.20$&        25.514  &\textbf{10.789} &        10.588  &        10.120       
                                         &        25.566  &\textbf{10.854} &        10.514  &        10.013 \\ 
    $^{13}_{~\Lambda}$C   & $12.0\pm0.2$ &        28.080  &\textbf{12.262} &        12.023  &        11.422  
                                         &        27.814  &\textbf{11.932} &        11.534  &        10.944 \\  
    $^{16}_{~\Lambda}$O   & $13.0\pm0.2$ &        27.516  &\textbf{12.849} &\textbf{12.643} &        12.174  
                                         &        27.771  &\textbf{13.089} &\textbf{12.684} &        12.105 \\
    $^{28}_{~\Lambda}$Si  & $17.2\pm0.2$ &        34.724  &\textbf{17.643} &\textbf{17.560} &        17.450  
                                         &        34.857  &\textbf{17.731} &\textbf{17.578} &        17.469 \\ 
    $^{32}_{~\Lambda}$S   & $17.5\pm0.5$ &        36.814  &\textbf{18.865} &\textbf{18.895} &        18.827  
                                         &        36.341  &\textbf{18.824} & \textbf{18.695}&        18.640 \\  
    $^{40}_{~\Lambda}$Ca  & $18.7\pm1.1$ &        36.600  &\textbf{19.566} &\textbf{19.448} &\textbf{19.265} 
                                         &        36.730  &\textbf{19.756} &\textbf{19.493} &\textbf{19.235} \\
    $^{51}_{~\Lambda}$V   & $21.5\pm0.6$ &        39.126  &\textbf{21.227} &\textbf{21.228} &\textbf{21.401} 
                                         &        39.095  &\textbf{21.251} &\textbf{21.221} &\textbf{21.407} \\ 
    $^{52}_{~\Lambda}$V   & $21.8\pm0.3$ &        39.429  &\textbf{21.422} &\textbf{21.440} &\textbf{21.662} 
                                         &        39.348  &\textbf{21.400} &\textbf{21.402} &\textbf{21.649} \\ 
    $^{89}_{~\Lambda}$Y   & $23.6\pm0.5$ &        41.882  &\textbf{23.511} &\textbf{23.576} &\textbf{23.974} 
                                         &        41.788  &\textbf{23.501} &\textbf{23.578} & \textbf{24.018} \\
    $^{139}_{~~\Lambda}$La& $25.1\pm1.2$ &        42.691  &\textbf{24.306} &\textbf{24.479} &\textbf{25.215} 
                                         &        42.226  &\textbf{23.987} &\textbf{24.295} &\textbf{25.210} \\
    $^{208}_{~~\Lambda}$Pb& $26.9\pm0.8$ &        44.489  &\textbf{25.687} &\textbf{25.893} &\textbf{26.746} 
                                         &        44.029  &\textbf{25.337} &\textbf{25.694} &\textbf{26.729} \\
    \hline
    $\bar{\chi}^2$      & {}             &                & 2.543          & 1.867          & 0.185          
                                         &                & 2.580          & 1.889          & 0.211           \\
    $\bar{\chi}^2_{\text{all}}$  
                        & {}             &     2956.907   & 2.543          & 3.009          & 6.690         
                                         &     2954.595   & 2.580          & 3.666          & 9.233           \\
    $\Delta$            & {}             &       17.175   & 0.711          & 0.672          & 0.668         
                                         &       17.048   & 0.816          & 0.712          & 0.716           \\
    $\delta$            & {}             &       98.221   & 3.759          & 3.840          & 4.817        
                                         &       97.776   & 4.018          & 4.060          & 5.419           \\  
    $R_\sigma$          & {}             & 0.667          & 0.366          & 0.417          & 0.577          
                                         & 0.667          & 0.367          & 0.464          & 0.659           \\
    $R_\omega$          & {}             & 0.667          & 0.352          & 0.415          & 0.611         
                                         & 0.667          & 0.353          & 0.472          & 0.712            \\ 
    $\sigma_{R_\sigma}^+$ & {}           & {}             & 0.053          & 0.036          & 0.080          
                                         & {}             & 0.080          & 0.092          & 0.084           \\
    $\sigma_{R_\sigma}^-$ & {}           & {}             & 0.079          & 0.071          & 0.082 
                                         & {}             & 0.088          & 0.093          & 0.085        \\ 
	\hline
	\hline
	\end{tabular}
}
\label{tab:DD-ME_fitting_result}
\end{table*}

Based on effective $NN$ interactions DD-ME2 and PKDD, 
the mass of $\Lambda$ hyperon is fixed to 1115.6 MeV and the two ratios $R_\sigma$ and 
$R_\omega$ are determined by minimizing the average square deviation
\begin{equation}
 \bar\chi^2(\bm{a}) = \frac{1}{N} \sum_i^N 
 \left(
     \dfrac{
	          B_{\Lambda,i}^{\text{exp.}} - B_{\Lambda,i}^{\text{cal.}}(x_i;\bm{a})
            }
            {\Delta B_{\Lambda,i}^{\text{exp.}}}
 \right)^2,
\label{eq:chi-square}
\end{equation}
where $\bm{a}$ is the ensemble of parameters to be determined ($R_\sigma$ and 
$R_\omega$), $i$ numbers each hypernucleus, $B_{\Lambda}$ is the separation 
energy and $\Delta B_{\Lambda}^{\text{exp.}}$ represents the experimental uncertainty. 
The experimental values of $B_{\Lambda,i}^{\text{exp.}}$ and 
$\Delta B_{\Lambda,i}^{\text{exp.}}$ are taken from 
Ref.~\cite{Gal2016_RMP88-035004} and references therein,
except for $^{40}_{~\Lambda}$Ca of which $B_{\Lambda,i}^{\text{exp.}}$ and 
$\Delta B_{\Lambda,i}^{\text{exp.}}$ are taken from Ref.~\cite{Pile1991_PRL66-2585}. 
The eleven hypernuclei cover a large mass interval with $A = 12$--208. 
When we made the fitting, it was found that the two ratios $R_\sigma$ and $R_\omega$
both deviate from 2/3: 
The more light hypernuclei are included in the fitting, the smaller the two ratios are.
This indicates that the in-medium $\Lambda N$ couplings are suppressed by structure 
effects in light hypernuclei.
To show such dependence of the deviations on the mass interval, we arrange these 
eleven hypernuclei into three groups: 
(1) All of them ($A = 12$--208), 
(2) nine of them ($^{16}_{~\Lambda}$O and heavier, i.e., $A = 16$--208) and 
(3) only six of them ($^{40}_{~\Lambda}$Ca and heavier ones, i.e., $A = 40$--208).
Based on either of the two effective $NN$ interactions DD-ME2 and PKDD, 
three new parameter sets are obtained. 
These six new effective interactions are labelled as DD-ME2-Y$i$ and PKDD-Y$i$ 
with $i=1$, 2 and 3 and listed in Table~\ref{tab:DD-ME_fitting_result}.

From Table~\ref{tab:DD-ME_fitting_result}, one can see that if only six
medium-heavy and heavy hypernuclei (those in Group 3) are used in the fitting, 
the average deviation $\bar\chi^2$ is the smallest which is around 0.2 
for both DD-ME2-Y3 and PKDD-Y3.
When more light hypernuclei are included, $\bar\chi^2$ becomes larger and
the two ratios $R_\sigma$ and $R_\omega$ become smaller. 
To check the overall description of the new effective interactions
for all of these eleven hypernuclei, we calculate and list in 
Table~\ref{tab:DD-ME_fitting_result} the average deviation 
$\bar\chi^2_{\rm all}$ as defined in Eq.~(\ref{eq:chi-square}),
the root mean square (rms) deviation $\Delta$ and 
the root of relative square (rrs) deviation $\delta$
\begin{equation}
\begin{split}
& \Delta = \sqrt{ \frac{1}{N} \sum_i^{N} 
                 \left(       { B_{\Lambda,i}^{\text{exp.}} - B_{\Lambda,i}^{\text{cal.}}}
                 \right)^2,
               }
 \\
& \delta = \sqrt{ \frac{1}{N} \sum_i^{N} 
                 \left( \dfrac{ B_{\Lambda,i}^{\text{exp.}} - B_{\Lambda,i}^{\text{cal.}}}
                              { B_{\Lambda,i}^{\text{exp.}}}
                 \right)^2,
               }
\label{eq:rrs}
\end{split}
\end{equation}
with $N = 11$, regardless how many hypernuclei are used in the fitting.
Therefore $\bar\chi^2_{\rm all}$ is larger than $\bar\chi^2$ for DD-ME2-Y$i$ and 
PKDD-Y$i$ with $i>1$.
For example, the average deviation $\bar\chi^2_{\rm all}$ is 6.690 for DD-ME2-Y3
and 9.233 for PKDD-Y3. 
Such large $\bar\chi^2_{\rm all}$ values are mainly due to the fact that 
the uncertainties of $\Lambda$ separation energies of light hypernuclei are quite small, 
$\Delta B_{\Lambda}^{\text{exp.}} = 0.2$ MeV 
for $^{12,13}_{~~~~\Lambda}$C, $^{16}_{~\Lambda}$O and $^{28}_{~\Lambda}$Si.
Compared with $\bar\chi^2_{\rm all}$, the rms and the rrs deviations are
more adequate in describing the agreement between the calculation and experiment:
$\Delta = 0.67$--0.82 MeV and $\delta = 3.8$--5.4\% which are fairly small, 
meaning a reasonably good agreement.

\begin{figure*}[htbp]
\centering
\includegraphics[width=0.9\textwidth]{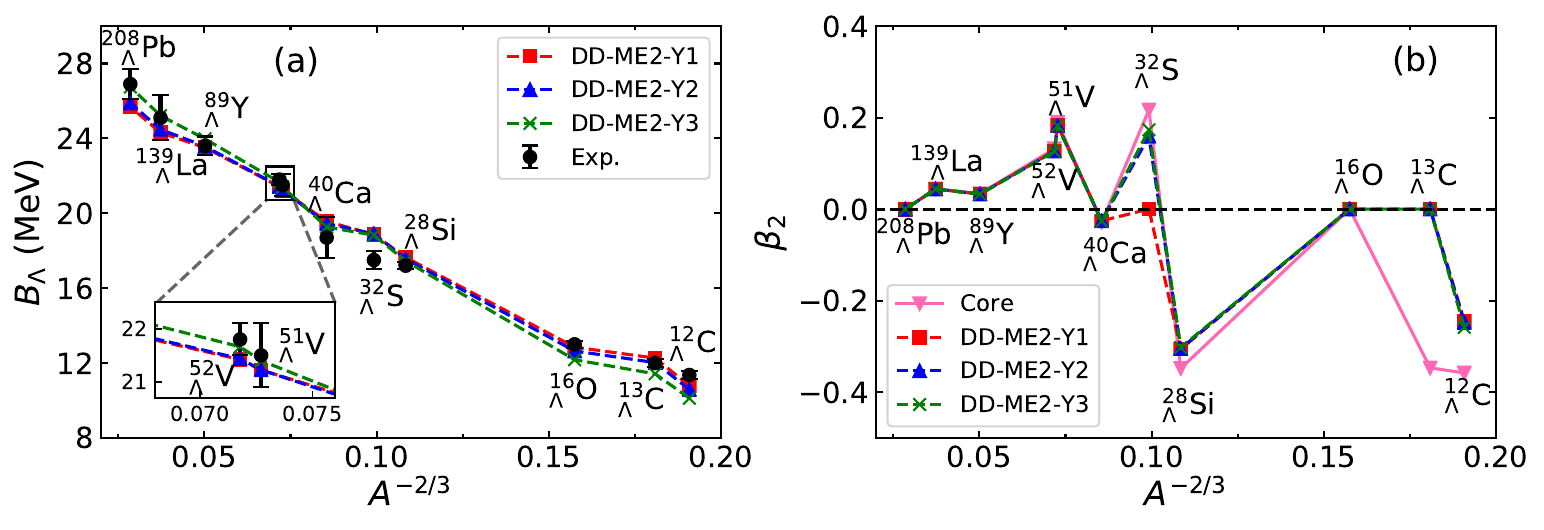}
\caption{(Color online) 
(a) The calculated $\Lambda$ separation energies compared with the experimental values and 
(b) the calculated quadrupole deformation parameters with DD-ME2-Y1, 
DD-ME2-Y2 and DD-ME2-Y3, respectively. 
The experimental values of $B_\Lambda$ are taken from~\cite{Gal2016_RMP88-035004} 
except $^{40}_\Lambda$Ca~\cite{Pile1991_PRL66-2585}. 
The inset in (a) shows the results for $^{51,52}_{~~~~\Lambda}$V.
The ``Core'' in (b) represents the quadrupole deformation parameters of 
the corresponding core nuclei calculated with DD-ME2. 
}
\label{fig:DD-ME2_BL_beta}
\end{figure*}


The $\Lambda$ separation energies $B_\Lambda$ calculated with DD-ME2-Y$i$ 
($i=1, 2$ and 3) are compared with experimental values in 
Fig.~\ref{fig:DD-ME2_BL_beta}(a).
It can be seen that each of these three new effective interactions can give 
a good description of $B_\Lambda$ for the selected hypernuclei 
with an exception of $^{32}_{~\Lambda}$S.
Although all of these eleven hypernuclei are used in the parametrization fitting
for DD-ME2-Y1, the light ones weigh more due to the smaller experimental errors 
$\Delta B_{\Lambda,i}^{\text{exp.}}$ (0.2 MeV).
Therefore, with DD-ME2-Y1, the calculated separation energies of the four 
light hypernuclei are very much close to the experimental values while those 
of heavier ones deviate more from the experiment.
With DD-ME2-Y3, the opposite is true: The calculated $B_\Lambda$ for 
heavy and medium heavy hypernuclei, which are used in the fitting, 
are very much close to the experiment while noticeble discrepancies can be seen 
for $^{12,13}_{~~~~\Lambda}$C and $^{16}_{~\Lambda}$O.
As for $^{32}_{~\Lambda}$S, the calculated $B_\Lambda$ with these three 
parameter sets are very similar (18.865, 18.895 and 18.827 MeV) and
all larger than the experimental value ($17.5 \pm 0.5$ MeV) considerably.
The reason for this discrepancy is not clear to us yet.

Fig.~\ref{fig:DD-ME2_BL_beta}(b) shows the quadrupole deformation parameters 
of the eleven selected hypernuclei calculated with DD-ME2-Y$i$ ($i=1,2$ and $3$) 
and of the corresponding core nuclei calculated with DD-ME2.
From Fig.~\ref{fig:DD-ME2_BL_beta}(b) one can find that 
the deformation parameter of a deformed hypernucleus is always smaller 
than that of its normal nuclear core. 
This is particularly true for light hypernuclei, e.g., $^{12}_{~\Lambda}$C.
Such shape polarization effects of $\Lambda$ have been discussed in 
Refs.~\cite{Hiyama1999_PRC59-2351,Zhou2007_PRC76-034312,Win2008_PRC78-054311,
Lu2011_PRC84-014328,Win2011_PRC83-014301,Isaka2011_PRC83-044323,Lu2014_PRC89-044307}.
Although $^{12}$C is obviously oblate with $\beta_2=-0.347$,
$^{13}_{~\Lambda}$C is spherical with DD-ME2-Y$i$ ($i=1, 2$ and 3);
such a shape change is similar to that discussed in 
Ref.~\cite{Lu2011_PRC84-014328} where NL-RMF functionals were used.
$^{32}_{~\Lambda}$S is spherical with DD-ME2-Y1 though its core, $^{31}$S, 
is moderately deformed with $\beta_2=0.217$. 
DD-ME-Y2 and DD-ME-Y3 both predict a prolate $^{32}_{~\Lambda}$S with $\beta_2$
slightly smaller than that of $^{31}$S.
Similar discussions hold for $\Lambda$ separation energies and deformation 
parameters calculated with PKDD-Y$i$ ($i=1, 2$ and 3) and will not be repeated. 

\begin{figure}[!htbp]
\centering
\includegraphics[width=0.47\textwidth]{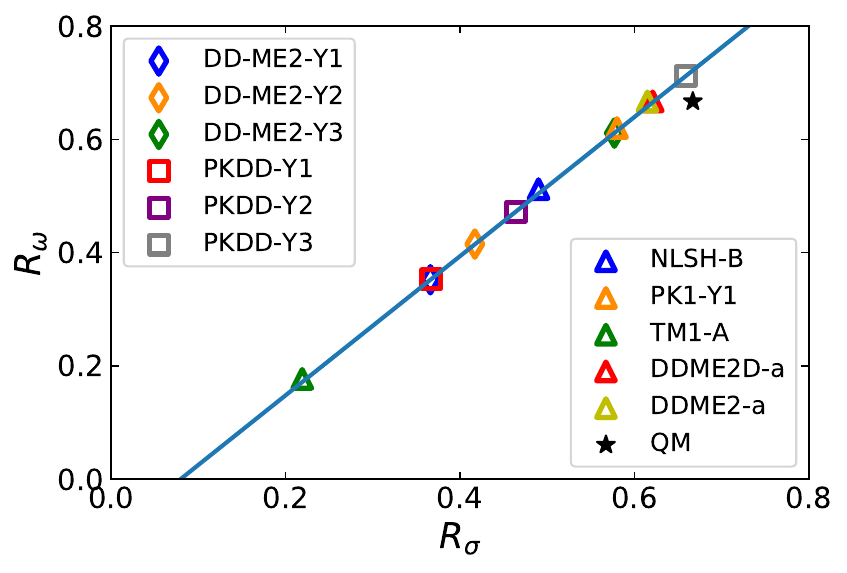}
\caption{(Color online) 
The correlation between $R_\sigma$ and $R_\omega$ in DD-ME2-Y$i$ and 
PKDD-Y$i$ ($i=1, 2$ and 3). 
The blue line shows the linear relation~(\ref{equ:linear_coupling-constant})
obtained by a linear fitting of $R_\sigma$ and $R_\omega$ in these six parameter sets.
Predictions from the quark model (QM), 
NL effective interactions NLSH-B~\cite{Ma1996_NPA608-305}, 
PK1-Y1~\cite{Wang2013_CTP60-479} and TM1-A~\cite{Sugahara1994_PTP92-803} and 
DD effective interactions DDME2D-a~\cite{Fortin2017_PRC95-065803} and 
DDME2-a~\cite{Fortin2017_PRC95-065803} are also shown for comparison.}
\label{fig:Rsig_Rome_relation}
\end{figure}

Among the six parameter sets DD-ME2-Y$i$ and PKDD-Y$i$ ($i=1, 2$ and 3), 
the two ratios $R_\sigma$ and $R_\omega$ change a lot. 
However, they are correlated linearly with each other as shown in 
Fig.~\ref{fig:Rsig_Rome_relation}. 
We made a linear fit of these two parameters and the relation 
\begin{equation}
 \label{equ:linear_coupling-constant}
 R_\omega = 1.228R_\sigma - 0.097,
\end{equation} 
is obtained and shown as the blue line in Fig.~\ref{fig:Rsig_Rome_relation}. 
This linear relation can be explained as follows.
In the RMF model, the central potentials $U_B$ ($B=N$ or $\Lambda$) for 
$\Lambda$ hypernuclei can be calculated from scalar and vector potentials approximately, 
$-U_B\approx g_{\sigma B}\sigma+g_{\omega B}\omega_0$. 
With the restriction $R_m=g_{m\Lambda}/g_{m N}$ ($m=\sigma$ or $\omega$), 
one obtains 
\begin{equation}
\label{equ:linear_para_from_potential}
 R_\omega \approx \dfrac{-U_\Lambda-R_\sigma g_{\sigma N}\sigma}{-U_N-g_{\sigma N}\sigma}.
\end{equation}
%
Generally speaking, the potential depths, represented by 
$U_B(0) \equiv U_B(r)|_{r=0}$, are about 70 MeV for nucleons and 30 MeV for $\Lambda$. 
The scalar potential depth $g_{\sigma N}\sigma(0)$ for nucleons is about $-400$ MeV. 
With these values, Eq.~(\ref{equ:linear_para_from_potential}) becomes 
$R_\omega\approx 1.212R_\sigma-0.091$, 
which is very close to Eq.~(\ref{equ:linear_coupling-constant}).
Similar linear behaviors between $R_\sigma$ and $R_\omega$ in nonlinear parameter
sets have been discussed in Refs.~\cite{Keil2000_PRC61-064309,Wang2013_CTP60-479}.
Several NL-RMF effective interactions NLSH-B~\cite{Ma1996_NPA608-305}, 
PK1-Y1~\cite{Wang2013_CTP60-479} and TM1-A~\cite{Sugahara1994_PTP92-803} and 
DD-RMF effective interactions
DDME2D-a~\cite{Fortin2017_PRC95-065803} and DDME2-a~\cite{Fortin2017_PRC95-065803},
are also shown in Fig.~\ref{fig:Rsig_Rome_relation}. 
They all fall well on the line defined in Eq.~(\ref{equ:linear_coupling-constant}).
The ratios $R_\sigma=R_\omega=2/3$ predicted from the quark model deviate from 
the blue line only slightly, as seen in Fig.~\ref{fig:Rsig_Rome_relation}. 
However, the $\Lambda$ separation energies calculated with the corresponding 
parameter sets (DD-ME2-Y0 and PKDD-Y0) are much larger than the experimental values 
as listed in Table~\ref{tab:DD-ME_fitting_result}. 
This means that these two ratios are connected strongly and correlated closely 
through the linear relation~(\ref{equ:linear_coupling-constant}). 
 
Next we analyze the errors of the parameters associated with the least-squares fitting 
by using the well known strategy for error estimates from statistical analysis~\cite{Brandt2014_Data-Analysis,Dobaczewski2014_JPG41-074001}.
For each effective interaction, a physically reasonable parameter space is defined 
by a confidence region around $R_\sigma$ and $R_\omega$ after normalization 
and the boundary of this space determines the errors of the parameters. 
Since the $\Lambda$ separation energy is a highly nonlinear function of the parameters, 
the obtained confidence region is asymmetric with respect to $R_\sigma$ and $R_\omega$. 
Given a certain value of $R_\sigma$, the error of $R_\omega$ is quite small 
(less than 0.001) due to the strong correlation between the two ratios 
[cf. Eq.~(\ref{equ:linear_coupling-constant})]. 
Therefore, we only evaluate the errors $\sigma_{R_\sigma}^+$ and $\sigma_{R_\sigma}^-$ 
of the independent parameter $R_{\sigma}$ for each effective interaction.
As seen in Table~\ref{tab:DD-ME_fitting_result}, $\sigma_{R_\sigma}^+$ and 
$\sigma_{R_\sigma}^-$ are smaller than 0.1 for all new effective interactions
proposed in this work.


\begin{figure*}[!htbp]
	\centering
	\includegraphics[width=0.8\textwidth]{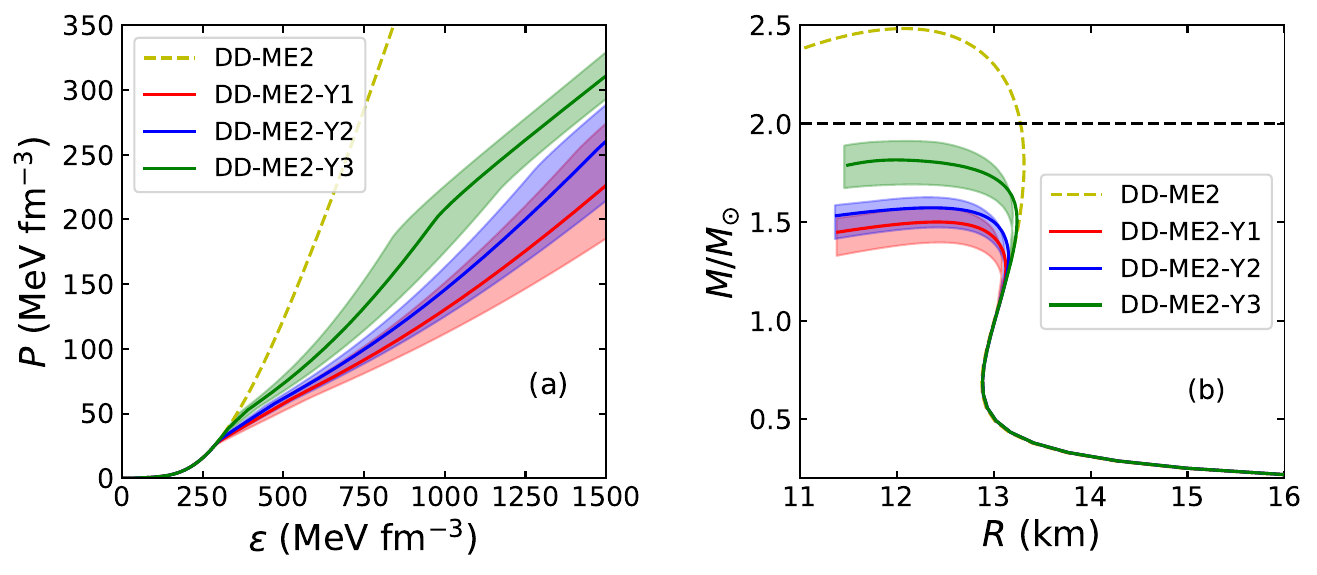}
	\caption{(Color online) 
		(a) The equations of state and (b) the mass--radius relations for neutron stars 
		with hyperons. 
		The solid curves are calculated with DD-ME2-Y$i$ ($i=1,2$ and 3) and the shaded 
		areas are calculated with the corresponding upper and lower boundaries of 
		$R_\sigma$ in the correlated direction defined by Eq.~(\ref{equ:linear_coupling-constant}). 
		Results without hyperons (calculated with DD-ME2) are also shown for comparison. 
	}
	\label{fig:DD-ME2_MR&EOS}
\end{figure*}

These new effective interactions are obtained by adjustments to properties of hypernuclei. 
The question then arises as to how well the neutron star properties can be described by them. 
The equation of state (EoS) and mass--radius ($M$--$R$) relation of neutron stars 
are calculated with DD-ME2-Y$i$ ($i=1,2$ and 3) and shown in Fig.~\ref{fig:DD-ME2_MR&EOS}. 
The octet baryons $p,n,\Lambda,\Sigma^{\pm},\Sigma^0, \Xi^0$ and $\Xi^-$ and 
the leptons $e^-$ and $\mu^-$ are considered. 
The vector coupling constants are determined by the naive quark model, i.e.,  
$2g_{\omega\Xi}=g_{\omega\Sigma}=2g_{\omega N}/3$ 
for $\omega Y$ coupling constants and 
$g_{\rho\Sigma}=2g_{\rho\Xi}=2g_{\rho N}$ 
for $\rho Y$ couplings. The scalar coupling constants $g_{\sigma\Sigma}$ and $g_{\sigma\Xi}$ are constrained by the empirical potentials 
$U_{\Sigma}^{(N)}=30$ 
MeV and  
$U_{\Xi}^{(N)}=-15$ MeV~\cite{Ishizuka2008_JPG35-085201}, respectively.
The EoSs calculated with DD-ME2-Y$i$ ($i=1,2$ and 3) are the same as that 
with DD-ME2 at low energy density region where only nucleons exist. 
When the energy density is larger than a certain value (about 300 MeV fm$^{-3}$), 
hyperons appear and the EoS is softer than that without hyperons,
leading to the so called ``hyperon puzzle''~\cite{Schulze2006_PRC73-058801,Vidana2013_NPA914-367}: Hyperons soften the EoS 
so that the maximum mass of neutron stars is smaller than 2$M_{\odot}$ 
which is the lower limit of the maximum neutron star mass as constrained from 
the astrophysical observations~\cite{Antoniadis2013_Science340-1233232,Cromartie2019_NatAstron4-72}.
It can be seen in Fig.~\ref{fig:DD-ME2_MR&EOS} that the larger 
the $R_{\sigma}$, the stiffer the EoS and the larger the maximum mass of neutron stars. 
With DD-ME2-Y1, DD-ME2-Y2 and DD-ME2-Y3, the maximum masses of neutron stars 
are, respectively, about 1.4$M_{\odot}$, 1.5$M_{\odot}$ and 1.8$M_{\odot}$ 
which are all smaller than 2.5$M_{\odot}$ with DD-ME2. 
The maximum mass calculated with the upper boundary of $R_\sigma$ in DD-ME2-Y3 
is 1.9$M_{\odot}$ which is still smaller than 2$M_{\odot}$.
One way to stiffen the EoS and thus increase the maximum mass of hyperon stars is 
to introduce an additional repulsion from the exchange of $\phi$ mesons 
in the RMF framework~\cite{Bednarek2012_AA543-A157}.
A systematic study of $\phi$-meson effects on the properties of hyperon stars 
in the DD-RMF model has been carried out and it was found that 
the 2$M_{\odot}$ limit of the maximum mass 
can be reached by using several relativistic density functionals with
the $\phi$ meson included~\cite{Tu2021_in-prep}.

\section{Summary}\label{sec4}

We investigate the effective interactions for $\Lambda$ hypernuclei in 
the density dependent relativistic mean field model and propose new parameter sets. 
Based on effective $NN$ interactions DD-ME2 and PKDD, the two ratios of 
scalar and vector coupling constants between effective $\Lambda N$ and 
$NN$ interactions, namely, $R_\sigma$ and $R_\omega$, are optimized by 
fitting calculated $\Lambda$ separation energies to experimental values 
of eleven single-$\Lambda$ hypernuclei with $A \ge 12$. 
The calculations were carried out by using the MDC-RMF model in which 
deformations are allowed for these hypernuclei and their normal core nuclei. 
With three ways of grouping and including these eleven selected $\Lambda$ 
hypernuclei in the fitting, six new effective interactions DD-ME2-Y$i$ and 
PKDD-Y$i$ ($i=1,2$ and $3$) are obtained.
The two ratios $R_\sigma$ and $R_\omega$ in these six new effective interactions 
vary largely. 
But they are correlated well with each other and follow closely a linear relation.
The statistical error of the independent parameter $R_\sigma$ is estimated and 
the error bars are within 0.1.

Ground state properties of the eleven selected hypernuclei and the corresponding 
core nuclei are described well by the MDC-RMF model. 
For core nuclei, the calculated binding energies agree satisfactorily with 
the experiment and most of them are deformed. 
For hypernuclei, the calculated $\Lambda$ separation energies are close to 
the experimental values with a small average square deviation weighed 
by experimental uncertainties. 
Shape polarization effects of $\Lambda$ and the shape change of a hypernucleus 
compared to its core nucleus are obvious in the $A\leq 40$ mass region. 
The newly proposed effective interactions are also used to calculate the EoS of
hypernuclear matter and study the mass and radius of neutron stars.
It turns out that they fall into the ``swamp'' which is full of effective 
interactions connected with the well known ``hyperon puzzle'':
The lower limit of the maximum neutron star mass, i.e., 2$M_{\odot}$, cannot 
be reached because the EoS is not stiff enough when hyperons are considered.

\section*{Acnowledgements}

Helpful discussions with 
Johann Haidenbauer, Hoai Le, Andreas Nogga, Xiang-Xiang Sun and Kun Wang 
are gratefully acknowledged. 
We thank Xiang-Xiang Sun for reading the manuscript and valuable suggestions. 
This work has been supported by 
the National Key R\&D Program of China (Grant No. 2018YFA0404402), 
the National Natural Science Foundation of China (Grants 
No. 11525524, No. 12070131001, No. 12047503, and No. 11961141004),
the Key Research Program of Frontier Sciences of Chinese Academy of Sciences (Grant No. QYZDB-SSWSYS013) and
the Strategic Priority Research Program of Chinese Academy of Sciences (Grants No. XDB34010000 and No. XDPB15). 
The results described in this paper are obtained on 
the High-performance Computing Cluster of ITP-CAS and
the ScGrid of the Supercomputing Center, Computer Network Information Center of Chinese Academy of Sciences.

\bibliographystyle{apsrev4-2}
\bibliography{../../../Notes/bib/ref}

\end{document}